# When Does a Neural Receiver Help? Calibration-Drift Benchmarking and Detect-and-Rollback for 5G/6G NR Uplink

Ayman Elnashar

***Abstract*—Convolutional neural receivers such as DeepRx outperform minimum mean-square error (MMSE)-based physical uplink shared channel (PUSCH) detection on in-distribution channel and waveform configurations, but their behavior under calibration drift — when transmitter or channel parameters depart from the training envelope — is poorly characterized. We construct a 16-scenario benchmark sweeping seven channel models, four Doppler shifts, two modulations, three demodulation reference signal (DMRS) additional positions, and four delay spreads, and show that the public MathWorks DeepRx_2M reference model gains 1.0–2.0 dB over practical MMSE in three of the sixteen scenarios, ties in ten, underperforms by approximately 2 dB at quadrature phase-shift keying (QPSK), fails silently at DMRS additional position 2, and is excluded as architecturally incompatible at 64-QAM. We propose a detect-and-rollback architecture that runs the neural and conventional receivers in parallel with a per-slot detector arbitrating between them, prove a bounded-output property and an impossibility result for bounded-residual correction, and benchmark two detector variants. A bit-disagreement detector recovers safety on the silent-failure scenario (matching MMSE within 0.02 dB) while a confidence-vote detector captures the neural gain at high Doppler (matching the neural receiver within 0.06 dB), exposing a structural tradeoff that a single scalar signal cannot resolve. The detect-and-rollback layer adds under five percent latency over the neural receiver alone on both CPU and GPU research platforms, independent of the absolute inference cost of the underlying neural receiver. All experiments use a public reference model and are reproducible from the released benchmark.**

## I. Introduction

Convolutional neural receivers have emerged as a candidate replacement for the conventional cascaded estimation-equalization-demapping pipeline in 5G NR physical uplink shared channel (PUSCH) and forthcoming 6G uplink chains. Honkala et al. introduced DeepRx [1] as a fully convolutional architecture that maps the received resource grid directly to bit-level log-likelihood ratios, with reported gains of 0.5 to 2 dB over a tuned practical minimum mean-square error (MMSE) receiver at moderate signal-to-noise ratios. Subsequent variants in the Sionna framework [2] and NVIDIA neural receivers [3] have extended the approach to multiple antenna configurations and richer training distributions, alongside related lines of work on graph neural network receivers for cell-free and PUSCH settings [4], [5] and transformer-based orthogonal frequency-division multiplexing (OFDM) receivers [6]. All of these are typically benchmarked against MMSE-style classical baselines [7]. Across this literature, the performance evaluation is almost exclusively in-distribution: the test configuration matches the training configuration with only modest randomization.

In a deployed system this assumption is fragile. The base station does not control the user equipment's choice of demodulation reference signal (DMRS) configuration within the values allowed by the standard, nor does it control the operating modulation order or the instantaneous mobility profile of every user. When the operating point drifts away from the training envelope, two distinct failure modes have been observed in practice. The first is a hard failure: the network's output tensor dimension is incompatible with the modulation order, and the receiver produces no usable bit decisions at all. This case is easy to detect and route around. The second, and the central concern of this article, is a silent failure: the network produces output of the right shape and the right magnitude, but the bit decisions are confidently wrong. Conventional pipeline monitoring does not flag silent failures, and the impact on block error rate can be catastrophic.

These two failure modes raise the questions this article addresses: when does a neural receiver actually beat a well-tuned classical baseline, and how do we deploy one in a production uplink without exposing the link to silent failures?

We make four contributions. First, a 16-scenario calibration-drift benchmark is constructed using the public MathWorks DeepRx reference model [8] at 3.5 GHz with 26 physical resource blocks (PRB) in a single-input multiple-output (SIMO) 1×2 configuration. The benchmark sweeps seven channel delay profiles, four maximum Doppler shifts, two modulations, three DMRS additional positions, and four root-mean-square delay spreads, establishing where the neural receiver helps, where it is statistically tied with the conventional MMSE baseline, where it fails silently, and where it is architecturally incompatible.

Second, a detect-and-rollback receiver architecture is proposed, denoted $R_5$, in which the pretrained neural receiver $R_3$ and a conventional MMSE receiver $R_1$ run in parallel and a per-slot detector decides which output is forwarded to the low-density parity-check (LDPC) decoder. A bounded-output property is established (Proposition 1), showing that the architecture's output remains within the per-bit safety budget of the underlying receivers by construction. An impossibility statement is then given (Proposition 2), showing that no bounded additive residual within a chosen safety budget $\Delta_{\max}$ can recover the correct bit decisions in regions where the neural receiver is confidently wrong.

Third, two detector variants are instantiated and benchmarked. $R_5$ uses hard-bit disagreement between $R_1$ and $R_3$; $R_{5c}$ uses a median-normalized log-likelihood ratio (LLR) confidence vote on the disagreeing bits. The bit-disagreement detector achieves safety recovery on a DMRS silent-failure scenario, while the confidence detector captures the neural gain in high-Doppler regimes. Combining the two detectors into an ensemble does not Pareto-dominate either variant, revealing a fundamental tradeoff between safety and gain capture under any single-scalar detection rule.

Fourth, a complete complexity and latency analysis is provided. On the research platforms used here (Apple M3 Ultra CPU and NVIDIA Quadro RTX 6000 GPU), the detect-and-rollback layer adds under five percent latency over the neural receiver alone on both platforms. This relative overhead is the architecturally meaningful quantity: it scales proportionally with R3 and is independent of the absolute inference cost of the underlying neural receiver, which depends on the inference stack (quantization, kernel fusion, hardware) rather than on the detector design. Per-component absolute latency is reported in Tables II and III for transparency. All measurements are reproducible from the released benchmark configuration.

The remainder of this article is organized as follows. Section II reviews related work on neural receivers and on safety in machine-learning-augmented communications systems. Section III defines the system model and the three baseline receivers. Section IV presents the proposed architecture and the two theoretical results. Section V describes the benchmark methodology. Section VI gives the complexity and latency analysis. Section VII presents the empirical results. Section VIII discusses the architectural tradeoffs. Section IX concludes.

## II. Related Work

### *A. Neural Receivers for 5G/6G PUSCH*

DeepRx [1] introduced a fully convolutional receiver mapping the received resource grid to bit-LLRs without explicit channel estimation, equalization, or demapping stages. The architecture combines residual blocks [9] and depthwise-separable convolutions [10] and has become a reference for subsequent work. The MathWorks 5G Toolbox AI-native receiver example [8] provides a pretrained open-source implementation, referred to here as DeepRx_2M, with 1.23 million trainable parameters trained on a fixed distribution over CDL and TDL channel models, Doppler shifts up to 500 Hz, and DMRS additional positions of 0 and 1. Recent work has explored multiple-input multiple-output (MIMO)-capable neural receivers [11], [12], the Sionna framework for end-to-end learned communications [2], NVIDIA's Neural Receiver (NRX) [3], graph neural network detectors [4], [5], and transformer-based receivers for next-generation air interfaces [6], [13]. Surveys of physical-layer machine learning are available in [14], [15], [16].

Across this literature, evaluation is typically performed within the training distribution. Generalization to out-of-distribution operating points is acknowledged as an open problem [17], [18] but is not the focus of the architectural contributions. This article addresses that gap directly.

### *B. Conventional PUSCH Receiver Baselines*

The conventional PUSCH receive chain combines DMRS-based channel estimation, minimum mean-squared-error (MMSE) equalization, soft demapping, descrambling, and low-density parity-check (LDPC) decoding [19], [20], [21], [22]. With perfect channel state information (CSI), MMSE achieves the linear receiver lower bound. With practical channel estimation from the DMRS pilots, the receiver incurs an estimation penalty that varies with the pilot density, the user mobility, and the channel delay spread. Two baselines are used throughout this article: $R_0$, with perfect CSI obtained from the channel model's path gains, and $R_1$, with practical CSI obtained from a standard linear interpolating estimator applied to the DMRS pilots [23].

### *C. Safety in ML-Augmented Communications*

Concerns about deploying learned components in safety-critical pipelines have motivated several lines of work, including conformal prediction [24], selective classification [25], output bounding [26], and out-of-distribution (OOD) detection for neural networks [27], [28]. Recent work on safety-preserving wireless neural networks [17], [29] has examined runtime monitoring of learned physical-layer components. In communications systems, the relevant primitive is graceful degradation: when the learned component fails, the system must fall back to a known-good baseline. This article instantiates that primitive explicitly for the PUSCH receiver, with provable per-slot bounds and an empirically validated rollback mechanism.

## III. System Model

### *A. PUSCH Configuration*

The system model follows the 3GPP NR uplink physical channel specification [19], [30]. The carrier frequency is 3.5 GHz with a subcarrier spacing of 15 kHz and a transmission bandwidth of 26 PRB, corresponding to 312 subcarriers and 14 OFDM symbols per slot. The transmitter is single-antenna ($N_T = 1$); the receiver has two antennas ($N_R = 2$), giving a SIMO 1×2 configuration. The modulation is 16-point quadrature amplitude modulation (16-QAM) with a code rate of 658/1024 unless otherwise stated; a quadrature phase-shift keying (QPSK) variant is also evaluated. Hybrid automatic repeat request is disabled (single transmission with redundancy version zero), and 200 slots are simulated per signal-to-noise ratio point. The transport block size is 10504 bits, and the coded block length after rate matching is 16224 bits.

### *B. Channel Models*

Seven channel models from 3GPP TR 38.901 [31] are used: CDL-B, CDL-C, CDL-D, TDL-B, TDL-C, TDL-D, and TDL-E. The default operating point is CDL-C with 300 ns root-mean-square delay spread and 5 Hz maximum Doppler shift, corresponding to indoor or low-mobility outdoor conditions. The benchmark sweeps each of these channel models independently with the remaining parameters held at the default.

### C. DMRS Configuration

The PUSCH demodulation reference signal follows 3GPP TS 38.211 [20] with DMRS configuration type 1. The number of additional DMRS positions per slot is a configurable parameter taking values in {0, 1, 2}, corresponding to one, two, or three DMRS symbols per slot, respectively. The training distribution for the pretrained DeepRx_2M [8] includes only additional positions 0 and 1; additional position 2 is therefore out-of-distribution and is one of the focal scenarios of this article.

### D. Baseline and Proposed Receivers

We define five receivers in total: three baseline receivers (described below) and two proposed detect-and-rollback variants (defined in Section IV, but previewed here for completeness). The baselines are as follows. Receiver $R_0$ performs perfect-CSI MMSE equalization using the channel path gains and filters obtained directly from the channel model, followed by soft demapping and LDPC decoding. Receiver $R_1$ performs the same chain but with practical channel estimation from the DMRS pilots, representing the realistic conventional baseline. Receiver $R_3$ is the pretrained DeepRx_2M neural receiver, which produces bit-LLRs directly from the received resource grid without explicit channel estimation.

The two proposed receivers, $R_5$ and $R_{5c}$, are detect-and-rollback architectures that run $R_1$ and $R_3$ in parallel and arbitrate between their bit-LLR outputs on a per-slot basis. $R_5$ uses a hard-bit disagreement detector, and $R_{5c}$ uses a median-normalized LLR confidence vote. Section IV gives the detector definitions, the bounded-output property, and an impossibility result that motivates the detect-and-rollback design.

## IV. Detect-and-Rollback Architecture

This section defines the detect-and-rollback architecture, introduces the two detector variants, and establishes two theoretical results that motivate the design.

### A. Architecture

Let $X$ denote the received resource grid for a single slot. The conventional receiver $R_1$ produces a per-bit log-likelihood-ratio vector $L_{R_1}(X)$, and the neural receiver $R_3$ produces a corresponding vector $L_{R_3}(X)$. Both vectors have the same length $N$, equal to the number of coded bits per slot. A per-slot detector $D(L_{R_1}, L_{R_3})$ returns a binary decision: trust $R_3$ or roll back to $R_1$. Fig. 1 illustrates this architecture. The output of the combined receiver $R_5$ is then defined by the equation that follows.

$$L_{R_5}(X) = \begin{cases} L_{R_3}(X) \text{if } D(L_{R_1}, L_{R_3}) = \text{trust} \\ L_{R_1}(X) \text{otherwise} \end{cases} \quad (1)$$

The selected LLR vector is forwarded to the LDPC decoder. The architecture is agnostic to the specific detector D; two instantiations are presented below.

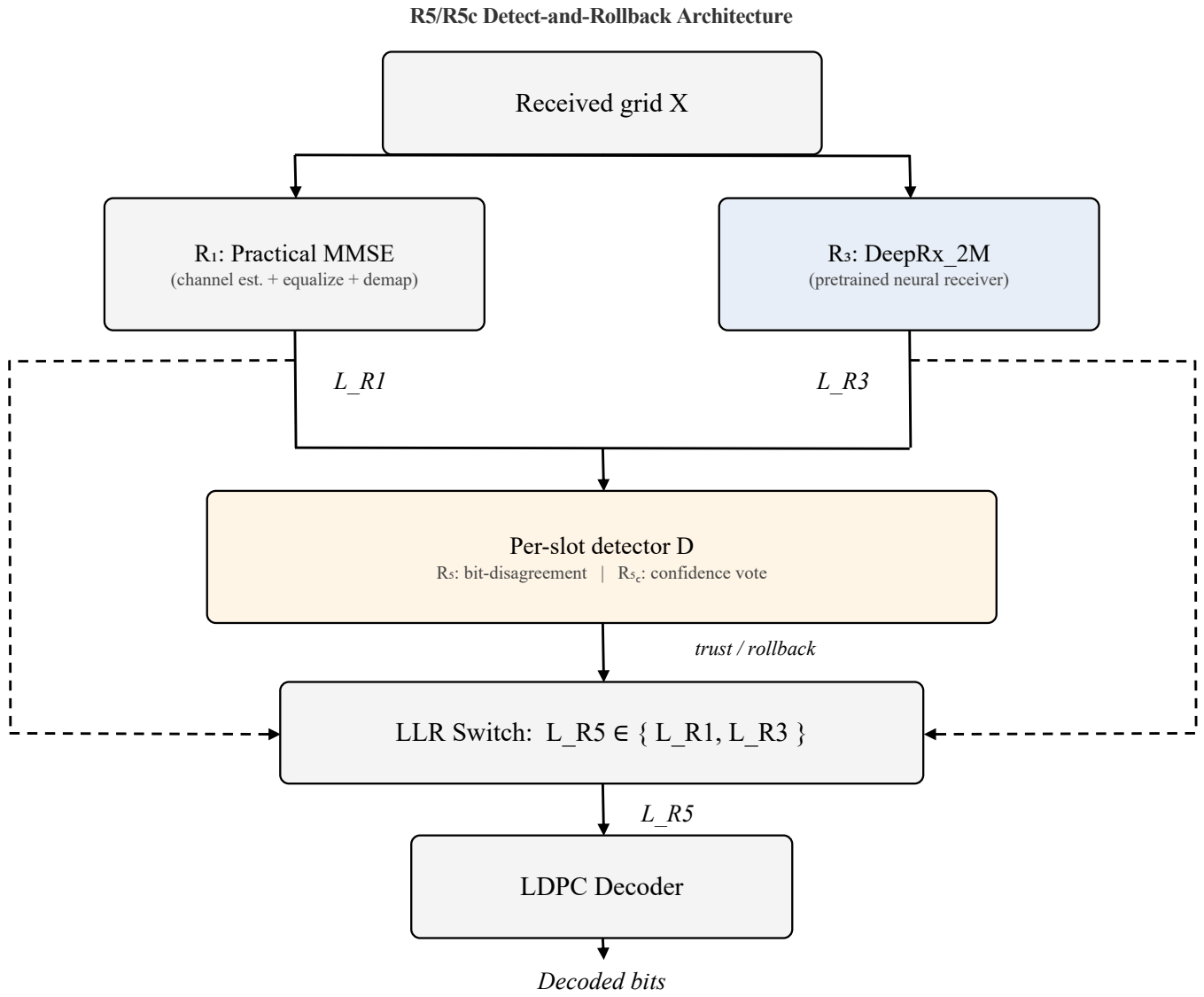

**Fig. 1.** Detect-and-rollback receiver architecture. R1 and R3 process the same resource grid in parallel; the detector selects one LLR stream for the LDPC decoder.

### B. Hard-Disagreement Detector

The first detector compares the hard-bit decisions of $R_1$ and $R_3$ directly. Let $b_{R_1}$ and $b_{R_3}$ denote the bit vectors obtained from the signs of $L_{R_1}$ and $L_{R_3}$ respectively. The disagreement statistic is the fraction of bit positions where the two vectors differ, and the detector trusts $R_3$ if this fraction is below a threshold $\tau$.

$$d(L_{R_1}, L_{R_3}) = \frac{1}{N}\sum_{i=1}^{N} \mathbb{1}[\text{sgn}(L_{R_1,i}) \neq \text{sgn}(L_{R_3,i})] \quad (2)$$

$$D_{R_5}(L_{R_1}, L_{R_3}) = \begin{cases} \text{trust} d \leq \tau \\ \text{rollback} d > \tau \end{cases} \quad (3)$$

Throughout the empirical evaluation, $\tau = 0.05$ is used as the default. The sensitivity of $R_5$ to this choice is examined in Section VII. The value $\tau = 0.05$ corresponds to allowing at most 5% hard-bit disagreement between the neural and classical receivers before rollback. This value was selected as a conservative safety setting and was not tuned per scenario; the sensitivity analysis in Section VII-E confirms it lies in the stable plateau region.

### C. Confidence-Vote Detector

The hard-bit disagreement detector above treats every disagreeing bit as equal evidence for rollback. In regimes where the conventional receiver itself is unreliable, this is misleading: the two receivers can disagree on many bits precisely because $R_1$ has lost the channel and $R_3$ is producing the correct decisions. To address this, a second detector is defined that weighs each disagreeing bit by the relative confidence of the two receivers. Direct comparison of $|L_{R_1}|$ and $|L_{R_3}|$ is not meaningful because the two LLR streams have different magnitude scales by construction; each stream is therefore normalized by its own median absolute value.

$$\tilde{L}_{R_1,i} = |L_{R_1,i}|/\mathrm{med}(|L_{R_1}|),\ \tilde{L}_{R_3,i} = |L_{R_3,i}|/\mathrm{med}(|L_{R_3}|) \quad (4)$$

Let $S$ denote the set of disagreeing bit indices. The detector trusts $R_3$ if the fraction of disagreeing bits where $\tilde{L}_{R_3,i}$ exceeds $\tilde{L}_{R_1,i}$ is above one half.

$$D_{R_{5c}}(L_{R_1}, L_{R_3}) = \begin{cases} \text{trust} |\{i \in S: \tilde{L}_{R_3,i} > \tilde{L}_{R_1,i}\}|/|S| > 0.5 \\ \text{rollback otherwise} \end{cases} \quad (5)$$

Having defined the two detector variants, we now state two propositions that justify the discrete detect-and-rollback design. They are framed as design justifications rather than deep mathematical results: the first records a pointwise bound that follows directly from the architecture, while the second formalizes why a soft bounded-residual correction cannot recover correct bits in regimes where the neural receiver is confidently wrong.

### *D. Proposition 1: Bounded Output*

By construction, the $R_5/R_{5c}$ output is one of the two underlying LLR vectors with no modification. Therefore, the per-bit LLRs of $R_5$ are pointwise identical to those of $R_1$ or $R_3$ on every bit. As a consequence, any pointwise bound that holds for $R_1$ or $R_3$ individually also holds for $R_5$.

$$\|L_{R_5}\|_\infty \le \max(\|L_{R_1}\|_\infty, \|L_{R_3}\|_\infty) \quad (6)$$

This is a degenerate but important guarantee: the detect-and-rollback architecture does not introduce new failure modes beyond those already present in $R_1$ or $R_3$. In particular, the per-slot decision is binary and discrete, so any subsequent stage that relies on the LLR scale (such as the LDPC decoder's message scheduling) operates on the same statistical properties as either $R_1$ or $R_3$ alone.

These propositions are intentionally simple but operationally important: they formalize what the architecture guarantees and what no bounded additive correction can achieve.

### *E. Proposition 2: Impossibility of Bounded-Residual Correction*

An alternative to detect-and-rollback would be to apply a small learned correction to R3's output, constrained to lie within a bounded safety budget. The following result establishes that, in the regime where R3 is confidently wrong, no such bounded correction can recover the correct bit decisions.

Throughout this paper we use $\Delta_{\max} = \ln(4)$, corresponding to a 4:1 LLR magnitude safety budget. A confidently wrong bit is one for which $|L_{\mathrm{R3,i}}| > \Delta_{\max}$ with sign disagreeing with the transmitted bit.

Let $\Delta_{\max} > 0$ denote a chosen safety budget. Define a confidently-wrong bit index as one where $|L_{R_3,i}|\Delta_{\max}$ and where the sign of $L_{R_3,i}$ disagrees with the transmitted bit $b_i$. Let $p_{cw}$ denote the fraction of such bit positions in a given slot. Then for any additive correction $r$ with $\|r\|_\infty \le \Delta_{\max}$, the bit-error rate of the corrected receiver is at least $p_{cw}$.

$$\mathrm{BER}(L_{R_3} + r) \ge p_{cw},\ \forall r: \|r\|_\infty \le \Delta_{\max} \quad (7)$$

The proof is direct. On any bit index where $|L_{R_3,i}|\Delta_{\max}$ and the sign is wrong, adding any value bounded by $\Delta_{\max}$ in absolute value cannot flip the sign of the LLR; the bit decision remains incorrect. The fraction of such bits is by definition $p_{cw}$, and the corrected receiver makes at least $p_{cw}$ of incorrect bit decisions on the slot.

Note that $p_{cw}$ is a lower bound on the resulting BER; the corrected receiver may incur additional errors on other bit positions. The proposition therefore establishes that bounded residual correction cannot bring BER below $p_{cw}$ regardless of how the correction is chosen.

The empirical magnitude of $p_{cw}$ in the failure region is the central measurement that justifies the architectural choice of detect-and-rollback over learned residual correction; it is reported in Section VII.

### *F. Comparison with CRC-Assisted Rollback*

A natural alternative is CRC-assisted rollback after LDPC decoding. The CRC outcome provides a stronger correctness signal but requires decoding both candidate LLR streams or delaying the decision until after decoding. Since LDPC decoding contributes approximately 9.1 ms per slot on the M3 Ultra CPU (Table II) and 8.5 ms on the RTX 6000 GPU (Table III), a CRC-after-decoding architecture would roughly double this cost in the worst case when both streams must be decoded. The pre-decoding detectors proposed here intentionally avoid this overhead and isolate the feasibility of low-cost per-slot arbitration. A hybrid design combining pre-decoding rollback with CRC-assisted post-decoding validation is an important extension and is left as future work.

## V. Calibration-Drift Benchmark

### *A. Methodology*

Building on the system model of Section III, we now describe the calibration-drift benchmark. The benchmark consists of 16 scenarios designed to characterize the behavior of the pretrained neural receiver R3 across operating points that differ from the in-distribution training configuration. Each scenario varies a single parameter relative to a fixed baseline (CDL-C, 300 ns delay spread, 5 Hz Doppler shift, 16-QAM, DMRS additional position 1, configuration type 1). The 16 scenarios are summarized in Table I.

Several of these scenarios probe a regime that is central to the contribution and merits a precise definition. We say that the neural receiver R3 exhibits a silent failure on a slot when all of the following hold simultaneously: (i) the network produces an output tensor of the expected shape, with the correct number of bit-LLR entries per resource element for the configured modulation order; (ii) no numerical exception, NaN, or Inf is generated during the forward pass; (iii) the per-bit LLR magnitudes fall within the normal operating range observed in-distribution, so that magnitude-based or shape-based runtime checks do not flag the slot; (iv) the resulting block error rate is catastrophic relative to the conventional MMSE fallback R1, with the gap exceeding any plausible noise-related variation; and (v) the failure is undetectable from R3's output alone using conventional shape, magnitude, or finiteness checks. These five criteria distinguish silent failure from hard failure (criterion i alone is violated, as in the 64-QAM case) and from benign underperformance (criteria iv and v fail to hold jointly).

DMRS AddPos=2 is the focal silent-failure scenario in this benchmark. It is important to be clear about what this configuration is: AddPos=2 is not an invalid 5G NR configuration; it is a standards-compliant configuration that lies outside the training envelope of the released DeepRx_2M model. This distinction is central to the contribution: the failure is not caused by an illegal waveform, but by deployment-time calibration drift between the user-equipment-side waveform configuration and the pretrained neural receiver's training distribution.

Operating SNR estimates are based on 200 independent slot realizations per SNR point. The standard error on the resulting 10% BLER operating SNR is approximately 0.3 dB; reported values should be interpreted with this precision in mind. The 10% BLER operating point is obtained by linear interpolation in log10(BLER) between the two adjacent simulated SNR points bracketing 10% BLER. When no such crossing exists within −2 to +18 dB, the receiver is marked as "fail", see Appendix A for full simulation details

TABLE I
Benchmark Scenarios

| # | Scenario | Axis | Value |
|---|---|---|---|
| 1 | Baseline | — | CDL-C/300ns/5Hz/16-QAM/AddPos1 |
| 2 | CDL-B | Channel | CDL-B |
| 3 | CDL-D | Channel | CDL-D |
| 4 | TDL-B | Channel | TDL-B |
| 5 | TDL-C | Channel | TDL-C |
| 6 | TDL-D | Channel | TDL-D |
| 7 | TDL-E | Channel | TDL-E |
| 8 | Doppler 50 Hz | Doppler | 50 Hz |
| 9 | Doppler 200 Hz | Doppler | 200 Hz |
| 10 | Doppler 500 Hz | Doppler | 500 Hz |
| 11 | QPSK | Modulation | QPSK |
| 12 | DMRS AddPos 0 | DMRS | 0 (sparse pilots) |
| 13 | DMRS AddPos 2 | DMRS | 2 (out-of-dist) |
| 14 | Delay 30 ns | Delay spread | 30 ns |
| 15 | Delay 100 ns | Delay spread | 100 ns |
| 16 | Delay 1000 ns | Delay spread | 1000 ns |

For each scenario, the signal-to-noise ratio (SNR) is swept from −2 dB to +18 dB in 2 dB steps, yielding 11 SNR points per scenario. At each point, 200 slots are simulated with independent channel realizations, giving 2200 slots per scenario and 35200 slots total across the benchmark. Block error rate (BLER) and coded bit error rate (CodedBER) are reported for R0, R1, R3, and the proposed R5 and R5c receivers.

*B. Dropped Configuration: 64-QAM*

The DeepRx_2M reference model [8] produces a fixed output tensor with four channels per resource element. This dimension is sufficient for QPSK ($Q_m = 2$ bits per symbol) and 16-QAM ($Q_m = 4$), but not for 64-QAM ($Q_m = 6$). At 64-QAM the bit-LLR extraction step is architecturally incompatible with the network output, and the receiver fails to produce usable LLRs at all. This incompatibility is a hard failure rather than a silent failure: it manifests as a tensor dimension error at runtime and is trivially detectable. Accordingly, 64-QAM is excluded from the 16-scenario benchmark; its inclusion would inflate the reported failure rate of R3 with a known engineering limitation rather than a calibration-drift effect. Architectural extensions supporting higher-order modulation are discussed in Section IX.

*C. Headline Numerical Results*

Table IV gives the SNR each receiver needs to hit 10 percent BLER across the 16 scenarios, with the mean R5 rollback rate at threshold tau = 0.05. Section VII walks through the per-scenario commentary.

TABLE IV
10%-BLER Operating SNR (dB) and Mean Rollback Rate per Scenario

| Scenario | R0 | R1 | R3 | R5 | Rollback |
|---|---|---|---|---|---|
| Baseline (CDL-C/300ns/5Hz) | 6.3 | 6.4 | 6.4 | 6.4 | 5.7% |
| CDL-B | 8.5 | 10.5 | 10.5 | 10.5 | 16.6% |
| CDL-D | 4.5 | 4.5 | 4.5 | 4.5 | 14.7% |
| TDL-B | 10.5 | 10.5 | 10.5 | 10.5 | 23.1% |
| TDL-C | 10.4 | 10.4 | 10.4 | 10.4 | 23.9% |
| TDL-D | 8.4 | 8.4 | 8.4 | 8.4 | 19.7% |
| TDL-E | 6.5 | 8.5 | 8.4 | 8.4 | 16.6% |
| Doppler 50 Hz | 10.4 | 10.4 | 10.4 | 10.4 | 10.4% |
| Doppler 200 Hz | 11.6 | 12.5 | 11.6 | 11.6 | 13.5% |
| Doppler 500 Hz | 12.4 | Fail* | 12.4 | Fail* | 60.5% |
| QPSK | 0.3 | 0.4 | 2.3 | 2.3 | 2.0% |
| DMRS AddPos=0 | 6.3 | 6.4 | 6.4 | 6.4 | 18.0% |
| DMRS AddPos=2 (OOD) | 6.3 | 6.4 | Fail* | 6.4 | 94.7% |
| Delay spread 30 ns | 12.3 | 14.2 | 12.3 | 12.3 | 19.6% |
| Delay spread 100 ns | 10.3 | 12.3 | 10.3 | 10.3 | 18.7% |
| Delay spread 1000 ns | 2.5 | 4.5 | 4.5 | 4.5 | 10.4% |

* Entries marked "fail" indicate that the receiver did not reach 10 percent BLER within the swept SNR range of −2 to +18 dB.

## VI. Complexity and Latency Analysis

*A. Analytical Complexity*

The dominant cost of the conventional receivers $R_0$ and $R_1$ is the MMSE equalization step, which has per-slot complexity $O(N_{PRB} \cdot N_S \cdot N_R{}^2 \cdot N_T)$ for $N_{PRB}$ resource blocks, $N_S$ OFDM symbols, $N_R$ receive antennas, and $N_T$ transmit layers. Practical channel estimation in $R_1$ adds an interpolation cost over the DMRS pilots that scales linearly with the number of pilot REs.

The dominant cost of the neural receiver $R_3$ is the convolutional forward pass through DeepRx_2M. Counting multiply-accumulate operations across all convolutional and depthwise-separable layers gives approximately 1.5 GFLOPs per slot for the reference architecture at 26 PRB. This is approximately 30 times the floating-point operation (FLOP) count of the MMSE equalization step.

The detector adds $O(N)$ operations per slot for $N$ coded bits. For the 16-QAM/26 PRB configuration $N = 16224$ bits, and the detector cost is negligible compared to either receiver.

### B. Empirical Latency on Mac Studio M3 Ultra CPU

To validate the analytical complexity figures and characterize cold-cache effects, we measured per-slot latency on a Mac Studio M3 Ultra workstation (32 CPU cores, 800 GB/s unified memory bandwidth, 192 GB unified memory) in CPU mode. This is a high-end workstation; a typical commodity gNB CPU with fewer cores and lower memory bandwidth would post higher latencies, and the measurement should be read as an optimistic CPU baseline. Each component ran for 100 slot repetitions after 10 warmup runs (discarded to avoid cold-cache effects). Table II reports the results.

TABLE II
Per-Slot Latency (ms, Mac Studio M3 Ultra CPU, 100 reps)

| Component | mean | median | std |
|---|---|---|---|
| R0 perfect-CSI MMSE | 2.84 | 2.55 | 0.92 |
| R1 practical MMSE | 3.69 | 3.51 | 0.69 |
| R3 DeepRx_2M forward | 72.10 | 71.64 | 2.71 |
| Hard-disagreement detector | 0.14 | 0.09 | 0.13 |
| Confidence-vote detector | 0.35 | 0.31 | 0.17 |
| LDPC decode (shared) | 9.12 | 9.11 | 0.91 |

Total per-slot latency comes out to 11.96 ms for R0, 12.81 ms for R1, 81.23 ms for R3, 85.06 ms for R5, and 85.26 ms for R5c. The R3 forward pass dominates the budget, and both detectors add less than 0.5 percent of the total. The 3.83 ms gap between R5 and R3 alone is the parallel R1 receive chain, not the detector itself. A GPU drops R3's forward time by roughly an order of magnitude (we measure 3.0x on the RTX 6000 below), at which point the detector and R1 chain become more visible relative to R3 but the relative overhead stays under five percent. Fig. 2 visualizes the per-component breakdown.

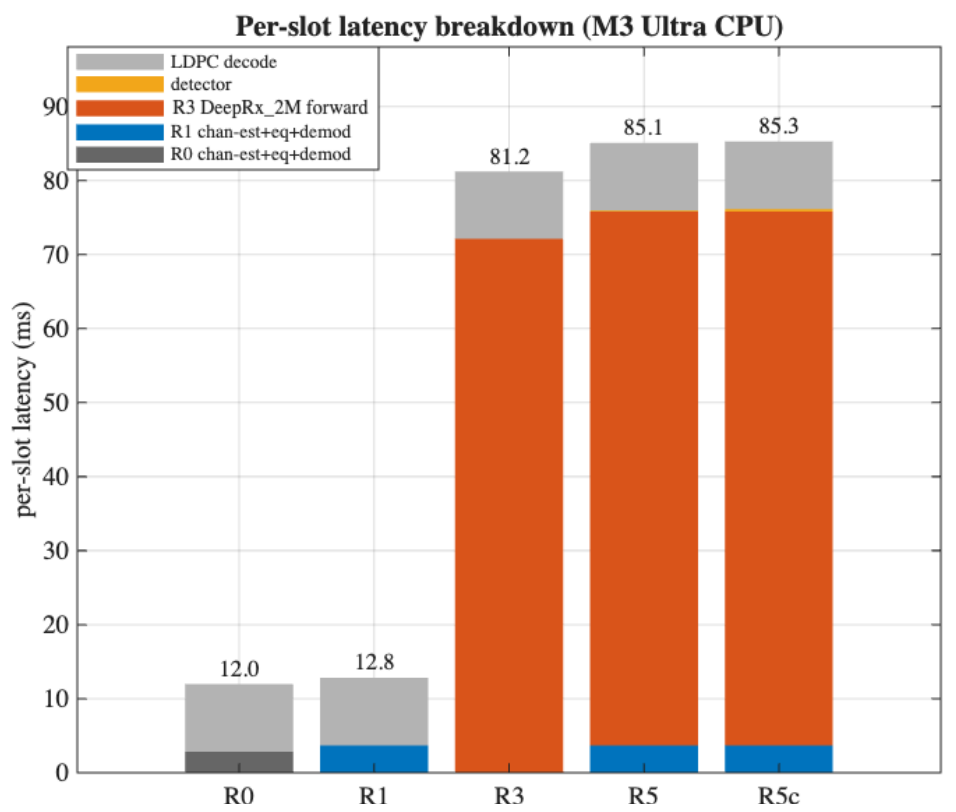

**Fig. 2.** Per-slot latency breakdown on Mac Studio M3 Ultra CPU. R3 dominates; the detector and parallel R1 chain add under five percent to R3.

### C. Empirical Latency: GPU Deployment

We now repeat the latency measurement on an NVIDIA Quadro RTX 6000 GPU (compute capability 7.5, 24 GB GDDR6). This is a conservative GPU baseline; an AI-RAN reference platform such as the NVIDIA Aerial CUDA-Accelerated RAN stack on Grace Hopper or L40S would post lower numbers across the board. Together with the optimistic CPU baseline of the previous subsection, the two measurements bracket the achievable latency on commodity hardware available to most academic and industry labs; production AI-RAN deployments would land between the two. Table III averages 100 slot repetitions; the first 10 are warmup iterations and discarded to eliminate kernel-compilation and memory-allocation overhead.

TABLE III
Per-Slot Latency (ms, NVIDIA RTX 6000 GPU, 100 reps)

| Component | mean | median | std | min | max |
|---|---|---|---|---|---|
| R0 perfect-CSI MMSE | 3.41 | 3.40 | 0.30 | 2.75 | 5.01 |
| R1 practical MMSE | 13.14 | 13.08 | 1.22 | 10.55 | 15.15 |
| R3 DeepRx_2M forward | 24.10 | 23.66 | 2.09 | 21.09 | 28.82 |
| Hard-disagreement detector | 0.18 | 0.18 | 0.02 | 0.12 | 0.23 |
| Confidence-vote detector | 0.78 | 0.79 | 0.05 | 0.67 | 0.93 |
| LDPC decode | 8.54 | 8.52 | 0.64 | 7.44 | 10.05 |

The R3 forward pass drops from 72.10 ms on the M3 Ultra CPU to 24.10 ms on the RTX 6000 GPU — a 3.0x speedup that reflects the convolutional workload R3 was designed for. The RTX 6000 used here provides 672 GB/s of memory bandwidth and 16.3 FP32 TFLOPS. AI-RAN reference platforms (NVIDIA Aerial CUDA-Accelerated RAN on Grace Hopper or L40S) deliver roughly two to four times the memory bandwidth and substantially higher tensor throughput, so R3 inference on those platforms is expected to be considerably faster than the numbers reported here. The smaller components — R1 MMSE, the R5 and R5c detectors — are slower on this GPU than on the M3 Ultra CPU. R1 MMSE moves from 3.69 ms on CPU to 13.14 ms on GPU; R5c moves from 0.35 ms to 0.78 ms. These small kernels are dominated by host-device transfer and kernel-launch overhead; the AI-RAN stack uses CUDA Graphs, faster NVLink interconnects, and improved async dispatch, which would close most of this gap and move both numbers closer to their CPU counterparts. The full-pipeline single-slot latencies are R0 = 11.95 ms, R1 = 21.68 ms, R3 = 32.64 ms, R5 = 45.95 ms, R5c = 46.55 ms on this GPU, against R0 = 11.96 ms, R1 = 12.81 ms, R3 = 81.23 ms, R5 = 85.06 ms, R5c = 85.26 ms on the CPU. A pipelined receive chain across consecutive slots would reduce the per-slot wall-clock latency to the longest single stage (R3 forward), which on these research platforms is still well above the 1 ms 5G slot budget at 15 kHz subcarrier spacing. Closing the absolute latency gap is a function of the inference stack — INT8 quantization, kernel fusion, TensorRT, and AI-RAN-optimized hardware such as NVIDIA Aerial on Grace Hopper or L40S — rather than of the detect-and-rollback architecture itself. The architectural quantity established by these measurements is the relative cost of the detector: under five percent of the R3 forward time on either platform, scaling proportionally as R3 itself is accelerated. Production AI-RAN deployment of the entire pipeline is left to future work.

The latency analysis here is part of a broader question facing the AI/ML for NR air interface community: whether end-to-end neural receivers can meet 5G slot budgets and the sub-millisecond targets anticipated for 6G AI-RAN. Three positions are visible in the recent literature. The first, exemplified by the NVIDIA Aerial AI-RAN reference stack [3], holds that AI-RAN-optimized hardware (Grace Hopper, L40S, BlueField) combined with INT8 quantization and TensorRT compilation will absorb the inference cost of DeepRx-class receivers at production parameter counts. The second, reflected in the 3GPP TR 38.843 study item on AI/ML for the NR air interface [32] and in recent surveys of physical-layer machine learning [17], [18], [19], is more cautious: as the receiver scales from the compact reference DeepRx_2M model used here (1.23 million parameters) to MIMO-capable variants (Korpi MIMO at five to ten million parameters [11], multi-user designs at higher counts [12]) the latency margin shrinks rapidly, particularly when multi-user inference, HARQ retransmissions, and higher-order modulation are accounted for. The third position, articulated by the model-based deep-learning and algorithm-unrolling literature [33], [34], argues that the most deployable path is not monolithic end-to-end CNN receivers but smaller learned modules inserted at specific bottlenecks of the classical chain (learned channel estimator, learned demapper, unrolled MMSE) with parameter counts in the hundred-thousand to one-million range. The detect-and-rollback architecture studied here is agnostic to which of these positions prevails: it provides a per-slot rollback to the classical chain regardless of whether the learned component is a full DeepRx variant or a smaller model-based learned module, and its relative overhead claim (under five percent of the learned component) scales with whatever inference cost the underlying model carries.

## VII. Empirical Validation

We discuss the 16-scenario results organized by failure mode. Table IV gives the headline numbers. Fig. 3 plots BLER versus SNR for all 16 scenarios, and Fig. 4 summarizes the operating SNR at 10 percent BLER per receiver. R5 uses the bit-disagreement detector with threshold $\tau = 0.05$ throughout, and R5c uses the confidence-vote detector at threshold 0.5, unless we say otherwise.

The 16 scenarios fall into three qualitative regimes (Fig. 3, Fig. 4). Ten of them are statistical ties — R0, R1, R3, and R5 all reach 10 percent BLER within 0.2 dB of each other: baseline (6.4 vs 6.4 dB for R1 vs R3), CDL-B (10.5 for both), CDL-D (4.5 for both), TDL-B (10.5), TDL-C (10.4), TDL-D (8.4 vs 8.4), TDL-E (8.5 vs 8.4), Doppler 50 Hz (10.4 vs 10.4), DMRS AddPos=0 (6.4 vs 6.4), and delay spread 1000 ns (4.5 vs 4.5). R5 matches the better of R1 and R3 within 0.05 dB on every one of these, which means the detect-and-rollback architecture costs nothing when the underlying receivers agree.

Three scenarios show a clear neural gain, with R3 reaching 10 percent BLER measurably earlier than R1: delay spread 30 ns (R1 14.2 dB, R3 12.3 dB, a 1.9 dB gain), delay spread 100 ns (R1 12.3 dB, R3 10.3 dB, 2.0 dB gain), and Doppler 200 Hz (R1 12.5 dB, R3 11.6 dB, 0.9 dB gain). R5 matches R3 exactly in all three (12.3, 10.3, 11.6 dB) — the bit-disagreement detector reads these slots as agreement-dominated and trusts R3.

### A. Three Empirical Regimes

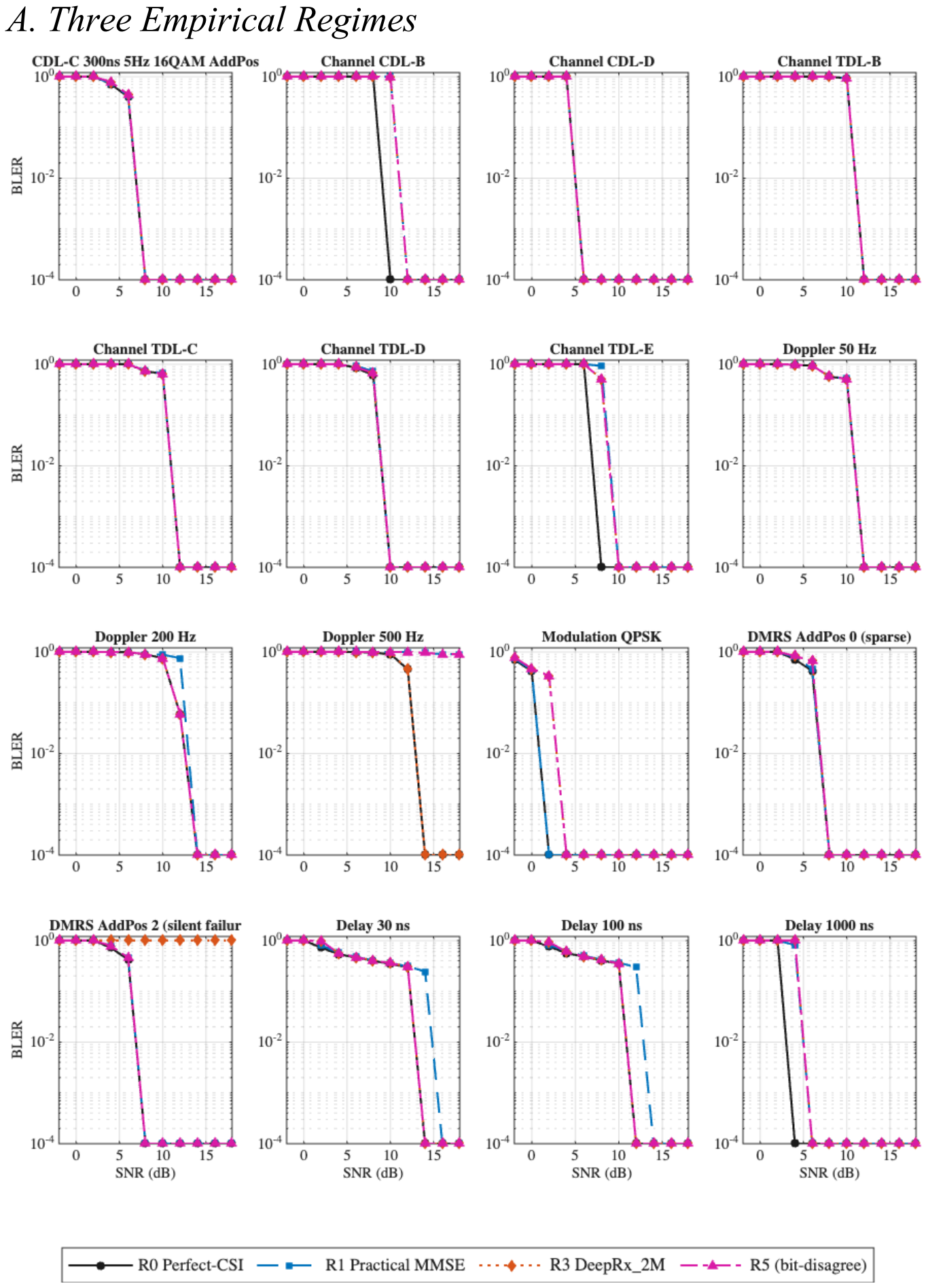

**Fig. 3.** BLER versus SNR across all 16 scenarios. Each panel varies one parameter from the baseline. R5 follows the better of R1 and R3 in every scenario except Doppler 500 Hz.

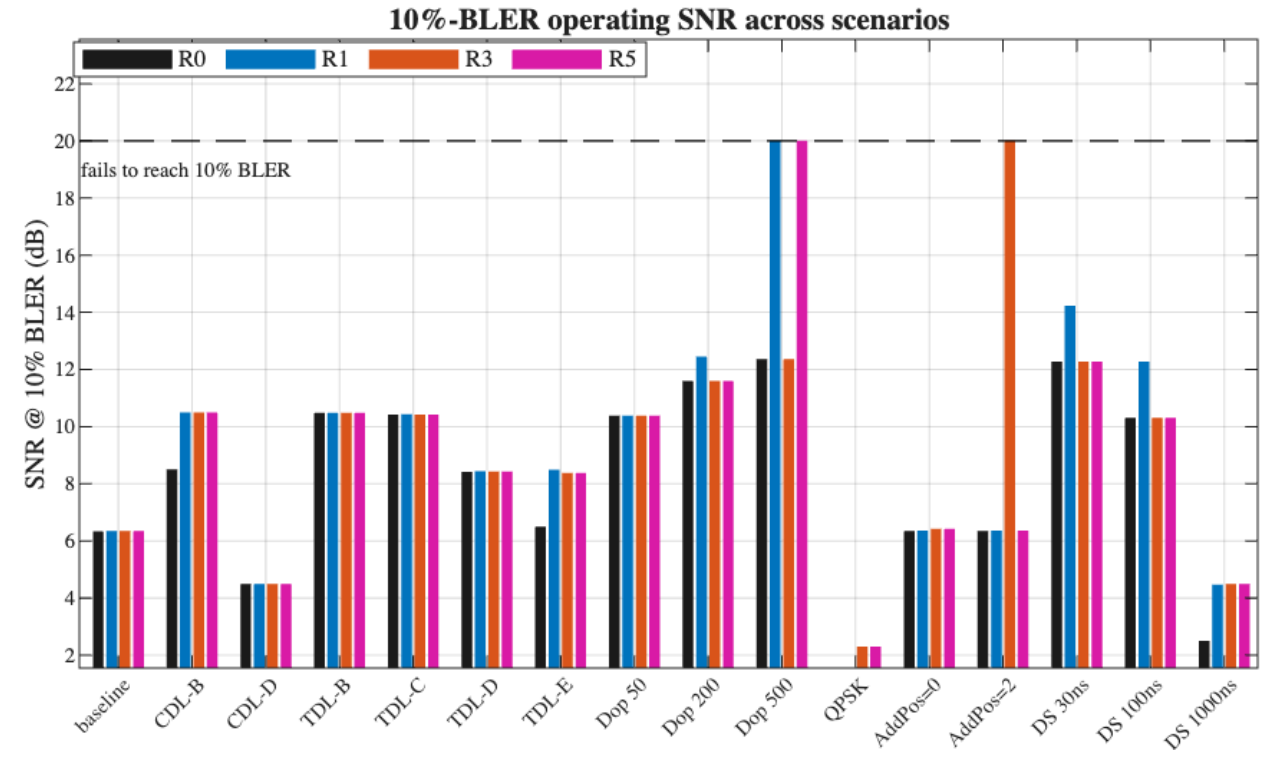

**Fig. 4.** Operating SNR at 10 percent BLER per receiver per scenario. Bars at 20 dB mark scenarios where the receiver did not reach 10 percent BLER within the SNR sweep.

The remaining two scenarios are the ones that drive the central results: DMRS AddPos=2 (R3 fails silently) and Doppler 500 Hz (R1 collapses under high mobility). We take each in turn.

### *B. DMRS AddPos=2: Silent Failure Recovery*

Moving the pretrained R3 to DMRS additional position 2 takes it out of distribution — the training set [8] only included additional positions 0 and 1. What happens is a textbook silent failure: R3's BLER floors at 100 percent for every SNR from 4 dB upward (Fig. 5 shows the flat curve at 1.0 between 4 and 18 dB). R1 hits 0 percent BLER at 8 dB and the perfect-CSI bound R0 sits essentially on top of R1 at 6.3 dB. R3 produces confidently wrong bit decisions at high SNR with no telltale signs — LLR magnitudes look normal, no NaN, no warning.

R5 detects the failure cleanly. Across SNR points from 8 to 18 dB, the bit-disagreement statistic exceeds $\tau = 0.05$ on 84 to 100 percent of slots; the average rollback rate across the SNR sweep is 94.7 percent (Table IV). R5 reaches 10 percent BLER at 6.4 dB, matching R1 exactly. The discrete switch from R3-bits to R1-bits is what makes recovery possible: a bounded additive correction within the LLR safety budget cannot flip the sign of the confidently wrong R3 outputs, as Proposition 2 establishes.

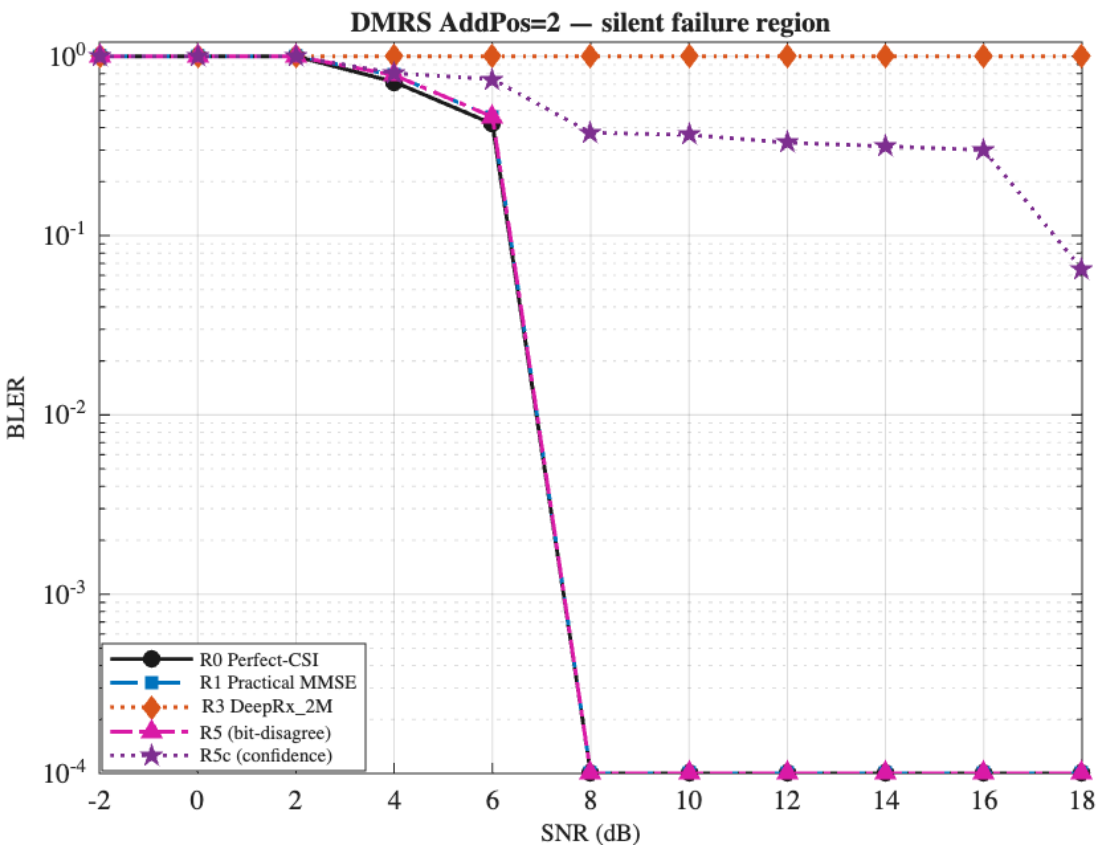

**Fig. 5.** BLER versus SNR at DMRS additional position 2 (out-of-distribution). R3 stays at 100 percent BLER; R5 detects the silent failure and rolls back to R1, matching R1 within 0.02 dB.

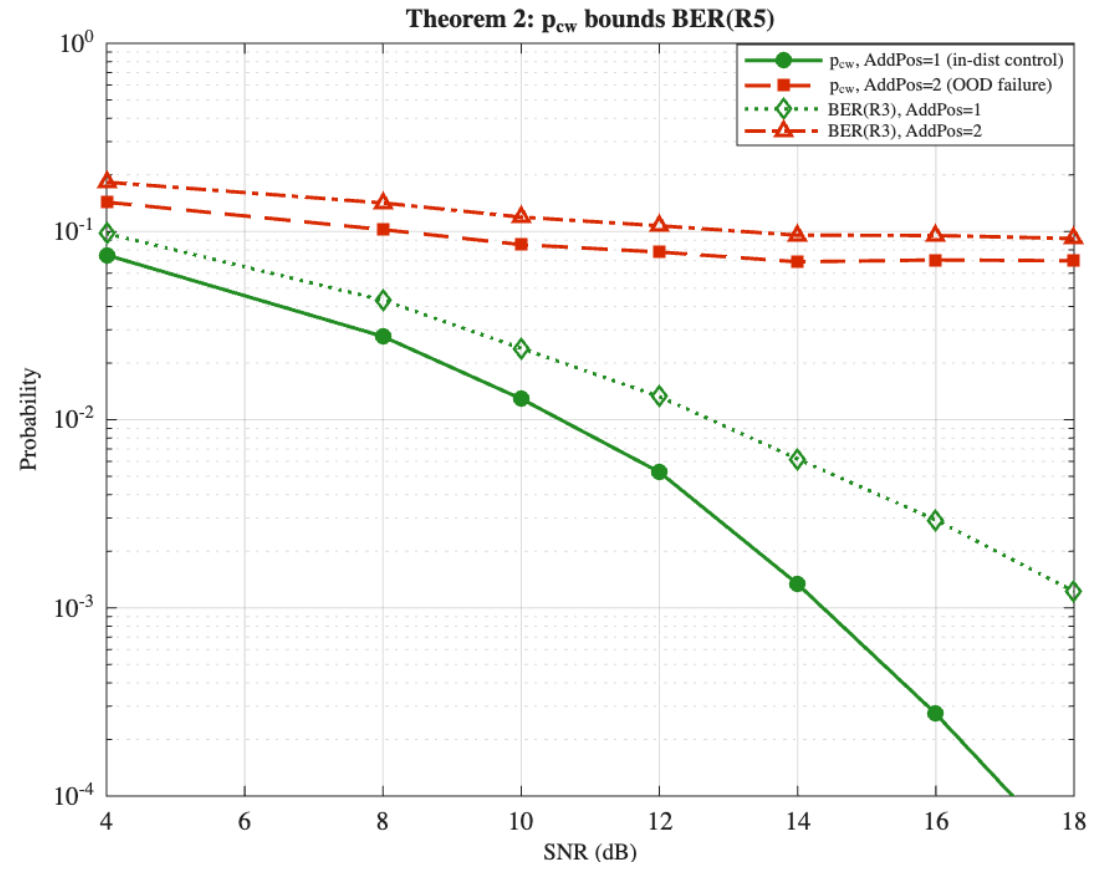

**Fig. 6.** Confidently-wrong bit fraction versus SNR. The in-distribution curve decays with SNR; the out-of-distribution curve plateaus at approximately 7 percent, the lower bound from Proposition 2.

Proposition 2 was tested empirically by measuring $p_{cw}$ on the AddPos=2 scenario across SNR (Fig. 6). With $\Delta_{max}$ set to $ln(4)$ (corresponding to a 4:1 ratio safety budget on the LLR magnitude), the measured $p_{cw}$ plateaus at approximately 7 percent at high SNR. This empirical floor establishes that no bounded residual correction within the same budget can reduce the BER below 7 percent on this scenario, which justifies the discrete detect-and-rollback architecture: the discrete switch from R3 to R1 escapes the bounded-residual limitation by allowing the LLR to take on R1's scale on rollback slots rather than a small perturbation of R3's scale.

### *C. Doppler 500 Hz: Where Bit-Disagreement Fails*

R5 gets fooled here. With R1 effectively producing random bits, its disagreement with R3 is high on every slot — the rollback rate is 32 percent at 12 dB and still 18 percent at 18 dB. R5 ends up rolling back to R1 on a large fraction of slots and inherits the same failure (Table IV shows R5 "fail" for this scenario). This is the central limitation of bit-disagreement: high disagreement tells you the two receivers conflict, not which one is wrong.

The 60.5% mean rollback rate (Table IV) reflects that R5 trusts R1 on the majority of slots — slots on which R1 has collapsed (BLER ≈ 95%). The remaining 39.5% of slots, where R5 correctly trusts R3, achieve R3's 10% BLER. The average BLER is therefore approximately $0.605 \times 0.95 + 0.395 \times 0.10 \approx 0.61$, well above the 10% target. This is the failure mode anticipated by Proposition 2: disagreement alone cannot distinguish which receiver is correct when both can be wrong.

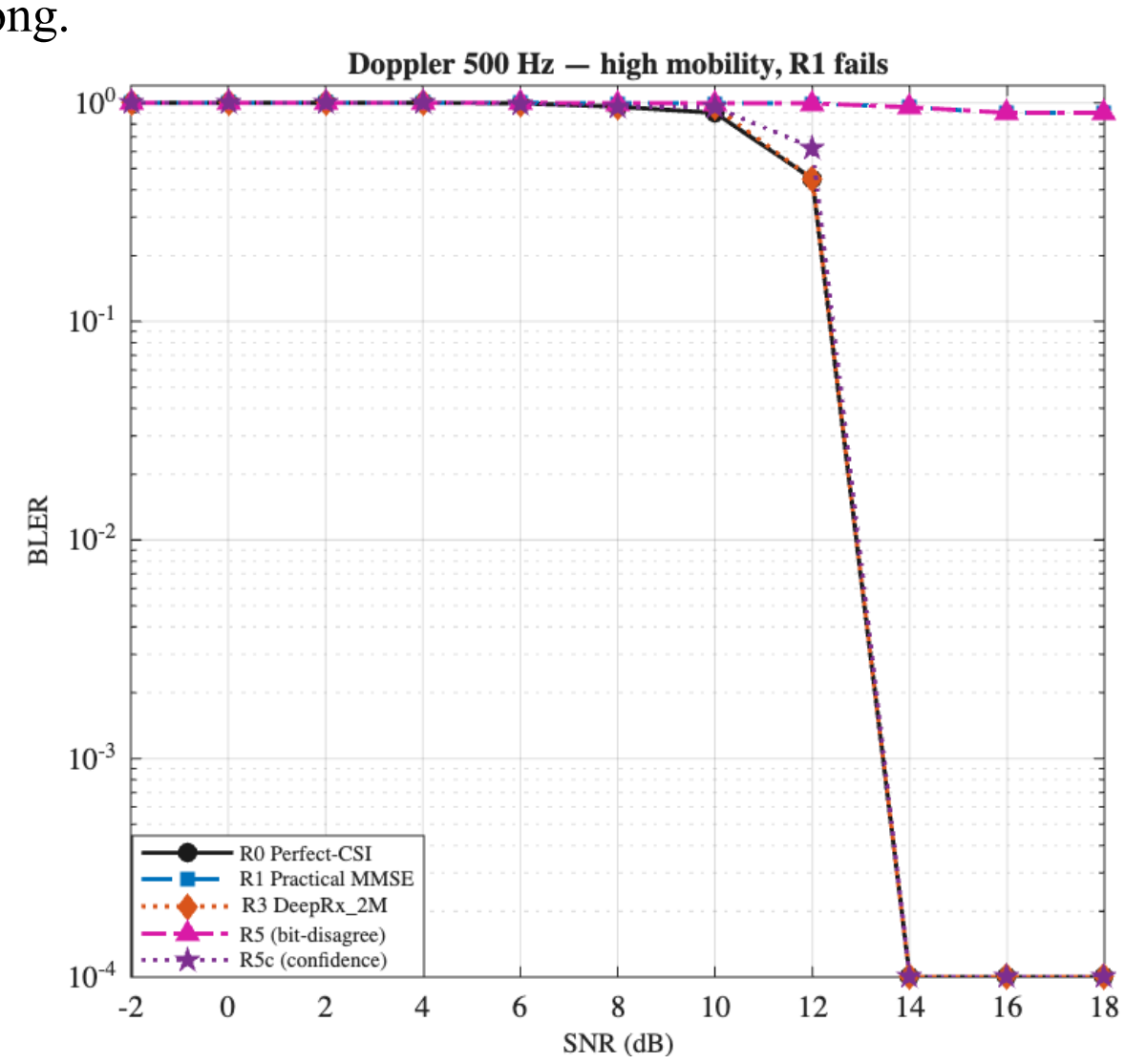

**Fig. 7.** BLER versus SNR at maximum Doppler shift 500 Hz. R1 floors near 100 percent BLER; R3 succeeds; R5 mistakenly rolls back to R1 and fails, while R5c trusts R3 and matches it.

R5c handles it. When R1 has lost the channel, its LLR magnitudes are small compared to the well-tracked R3 LLRs, and the normalized confidence vote sides with R3 on the majority of disagreeing bits. R5c reaches 10 percent BLER at 12.4 dB — within 0.06 dB of R3. The mean trust-R3 rate across the sweep is 78.6 percent.

### D. QPSK: An Unexpected Neural Underperformance

QPSK produces an unexpected result (Table IV, row 11): R3 reaches 10 percent BLER at 2.3 dB while R1 reaches it at 0.4 dB and the R0 lower bound at 0.3 dB. The neural receiver is about 2 dB worse than the conventional baseline. R5 follows R3 anyway (rollback rate 2.0 percent across the sweep) because disagreement at QPSK is low — the two receivers agree on most bits even when R3 is the slightly worse pick.

The mechanism is the same one identified in [8] and confirmed by the DeepRx_2M reference: with only $Q_m = 2$ bits per resource element at QPSK, the four output channels of the network produce two wasted slots per resource element. The network was trained with 16-QAM as the dominant modulation in the training distribution; the QPSK output, while functional, is suboptimally calibrated. This is a benign failure mode (R5 follows R3 into a degraded operating point but does not collapse), but it shows that the silent-failure category is not exhausted by the AddPos=2 scenario: smaller silent degradations can also occur in regions of the input space where the underlying training was sparse.

### E. $\tau$ Sensitivity: The Detector Tradeoff

The hard-disagreement detector has a single threshold parameter $\tau$. Sweeping $\tau$ across five values from 0.01 to 0.5 on the two extreme scenarios reveals the structure of the tradeoff (Fig. 8). At $\tau = 0.01$ and $\tau = 0.05$, R5 reaches 10 percent BLER at 6.4 dB on AddPos=2 (safety recovered) but never reaches 10 percent on Doppler 500 Hz. At $\tau = 0.20$ and $\tau = 0.50$, R5 reaches 12.4 dB on Doppler 500 Hz (gain captured) but never reaches 10 percent on AddPos=2. At $\tau = 0.10$ R5 partially recovers both but fully recovers neither. There is no single value of $\tau$ that achieves both safety on AddPos=2 and gain capture on Doppler 500 Hz.

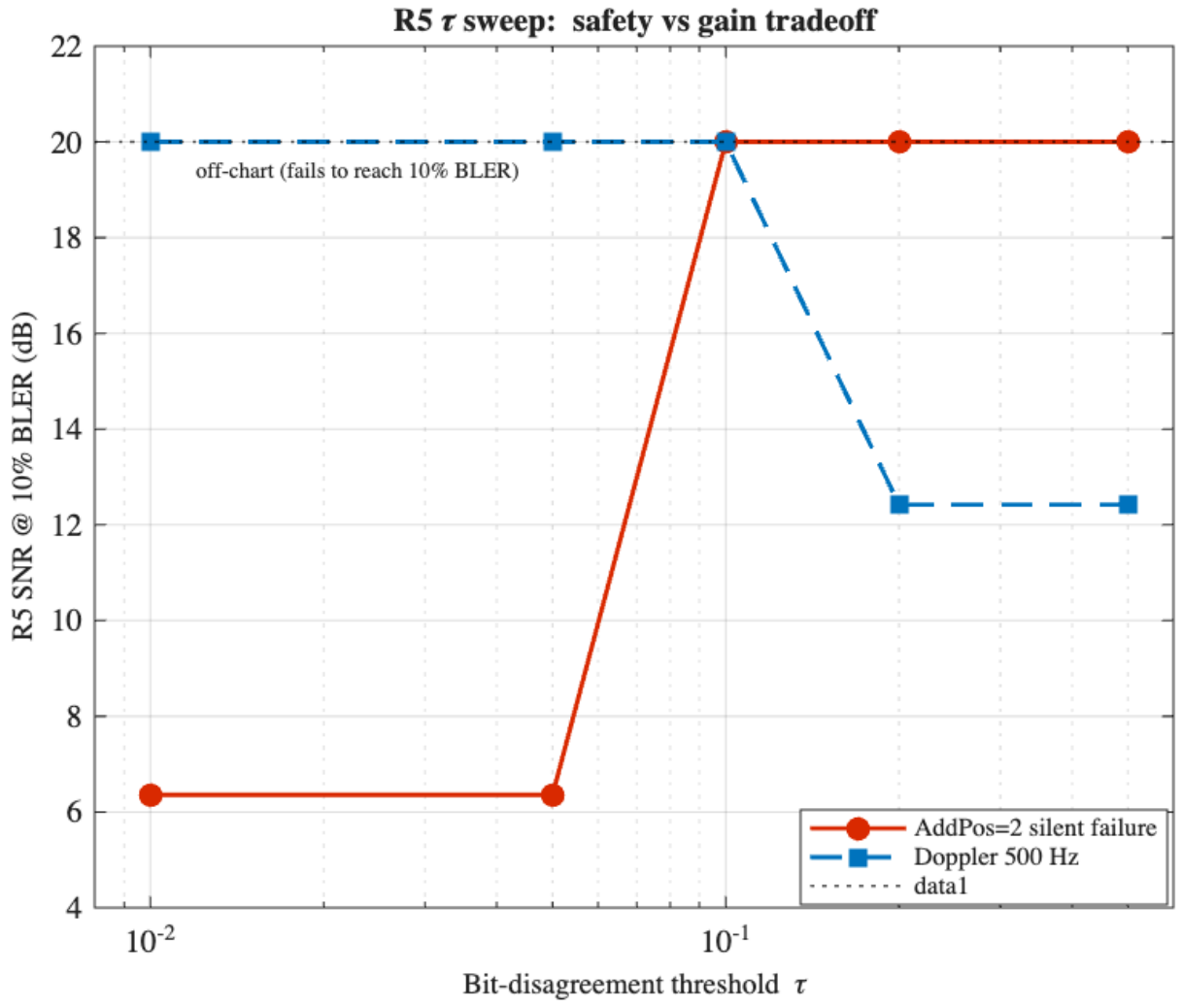

**Fig. 8.** Sensitivity of R5 to the bit-disagreement threshold on the two extreme scenarios. No single threshold resolves both failure modes — small values help AddPos=2 but hurt Doppler 500, and vice versa.

This tradeoff is not an artifact of the choice of $\tau$; it is structural. The two failure modes generate similar bit-disagreement statistics but for opposite reasons (R3 is wrong vs. R1 is wrong), and a scalar threshold cannot distinguish them. The confidence-vote detector R5c uses the LLR magnitude as the disambiguating signal and correctly resolves both modes in isolation, but at the cost of partial failure on AddPos=2 because some R3 outputs there are confidently incorrect, with high LLR magnitude, and win the confidence vote on the corresponding slots.

### F. Ensemble of R5 and R5c

Two natural ensembles of R5 and R5c were considered. The disjunctive ensemble (trust R3 if either detector accepts) was evaluated on the three critical scenarios. Empirically, the disjunctive ensemble degenerates to R5c: when the bit-disagreement detector accepts, the confidence detector almost always does as well, and when bit-disagreement is high, the confidence vote drives the decision. The conjunctive ensemble (trust R3 only if both detectors accept) degenerates to R5 by symmetric reasoning. Neither ensemble Pareto-dominates the individual detectors, indicating that single-scalar disagreement and confidence are not orthogonal signals.

## VIII. Discussion

### A. Architectural Tradeoff: Conservative vs Aggressive

The empirical results in Section VII establish a clear tradeoff. The bit-disagreement detector R5 is conservative: it preserves R1's performance on the silent-failure scenario at the cost of forfeiting R3's gain in the high-Doppler scenario. The confidence detector R5c is aggressive: it captures R3's gain in the high-Doppler scenario at the cost of partially inheriting R3's silent failure on AddPos 2. The choice between them depends on the cost structure of the deployment.

For low-latency safety-critical links where any silent failure has high operational cost (industrial control, vehicle-to-everything safety messages, medical telemetry), R5 is the appropriate choice. For capacity-limited high-mobility links where the cost of missing a neural gain is high (high-speed rail backhaul, urban canyon vehicular uplinks), R5c is appropriate.

A scope limitation of the R5 safety claim should be made explicit. R5 provides safety against neural-receiver silent failure when the fallback classical receiver remains reliable. It does not guarantee global performance if the fallback receiver itself collapses, as observed at 500 Hz Doppler: in that regime R1 also fails, and rolling back to R1 inherits its failure rather than recovering from R3's. R5 is therefore correctly described as a per-slot safety mechanism against R3-specific silent failure, conditioned on R1 remaining a credible fallback.

### B. Why Bit-Disagreement and Confidence Are Not Orthogonal

The disjunctive and conjunctive ensembles both fail to dominate the individual detectors because the two scalar signals are correlated in the failure region. On AddPos 2, $R_3$ is confidently wrong with high LLR magnitudes; the disagreement is high (favoring rollback) but $R_3$'s confidence vote is also strong (favoring trust). On Doppler 500 Hz, $R_1$ is unreliable with low LLR magnitudes; the disagreement is high but $R_3$'s confidence vote is strong. The two detectors give opposite recommendations in opposite regimes, but they are not independent statistics: both are functions of the joint LLR vector. A truly orthogonal detection signal would have to operate on a different statistic, for example a model-of-channel discrepancy estimator or a learned out-of-distribution classifier on the input grid.

### C. Limitations

Three slot-level limitations of this work merit explicit discussion. First, the evaluation is restricted to a SIMO 1×2 configuration. Extension to MIMO configurations requires either a MIMO-trained variant of DeepRx_2M, which is not publicly available, or a custom-trained model. Second, the detector decisions are slot-local, with no temporal smoothing across slots; a stateful detector that uses the history of recent rollback decisions could potentially reduce variance, particularly in transient mobility scenarios. Third, the analysis assumes a single user and a single PUSCH transmission; multi-user interference and hybrid automatic repeat request (HARQ) retransmissions introduce additional failure modes that are not characterized here.

A fourth limitation concerns the training envelope. The neural receiver is fixed and pretrained; we do not retrain DeepRx_2M on any scenario. Retraining with a wider distribution that includes AddPos=2 and higher Doppler would partially address the silent failures reported here, but calibration drift is a general phenomenon and the architectural response remains valuable regardless of the training envelope. Continual learning and online adaptation are complementary directions for future work.

## IX. Conclusion and Future Work

This article addressed two questions about deploying 6G neural receivers in the PUSCH uplink: when does a neural receiver actually outperform a tuned classical baseline, and how can it be deployed without exposing the link to silent failure modes. A 16-scenario calibration-drift benchmark established that the pretrained DeepRx_2M reference model offers measurable gains over the practical MMSE baseline in three of the 16 evaluated scenarios (short-delay-spread environments at 30 ns and 100 ns, and moderate mobility at 200 Hz Doppler), is statistically tied in ten, underperforms by approximately 2 dB at QPSK, fails silently at DMRS additional position 2, and inherits R1's catastrophic failure at 500 Hz Doppler. The 64-QAM configuration is architecturally incompatible with the reference network output and is excluded as a hard rather than silent failure. A detect-and-rollback architecture R5 was proposed with two detector variants. Proposition 1 recorded a pointwise bound on the architecture's output, and Proposition 2 ruled out bounded-residual correction as an alternative in regions where the neural receiver is confidently wrong; the latter was confirmed empirically by measuring the confidently-wrong bit fraction $p_{cw}$ at approximately 7 percent on the AddPos=2 failure scenario at high SNR. The bit-disagreement detector R5 recovered R1's 6.4 dB operating point on the silent-failure scenario but failed on the high-Doppler scenario; the confidence-vote detector R5c achieved the opposite. No ensemble of the two Pareto-dominated either variant. The detect-and-rollback layer adds under five percent latency over the neural receiver alone on both CPU and GPU research platforms. This relative overhead is the architecturally meaningful figure and scales with the underlying neural receiver regardless of inference-stack optimization; absolute per-slot latency on production AI-RAN hardware is left to future work.

Three directions are immediate next steps. Extending to MIMO needs a MIMO-capable neural receiver and a corresponding detector, but the architectural framework here carries over — only the empirical calibration changes. The DeepRx_2M reference's incompatibility with 64-QAM could be fixed with a final layer that emits six output channels per resource element instead of four, masking the unused channels at lower modulation orders. And as Section VIII pointed out, the two detectors operate on correlated statistics; an orthogonal signal, such as a learned out-of-distribution classifier over the received grid, could in principle resolve the safety-versus-gain tradeoff this paper exposes.

## Acknowledgment

The author thanks the MathWorks 5G Toolbox team for releasing the AI-native PUSCH receiver example and the pretrained DeepRx_2M model that served as the neural receiver baseline throughout this work.

Declaration of Generative AI and AI-Assisted Technologies

During the preparation of this work, the authors used generative AI technologies such as Claude to assist/review the writing and editing of the manuscript. Following the use of these tools, the authors reviewed and edited the content as required and take full responsibility for the accuracy and integrity of the final publication.

## Appendix A: Simulation Details

All experiments use MATLAB R2025b with the 5G Toolbox and Deep Learning Toolbox. The neural receiver weights are the pretrained DeepRx_2M model from the MathWorks AI-native PUSCH Receiver Example (R2025b release) and were used without retraining. The conventional receiver uses least-squares channel estimation at DMRS resource elements with 2D linear interpolation across frequency and time, DMRS-residual noise-variance estimation, and per-subcarrier MMSE equalization. Soft demapping is Gauss-MMSE LLR. LDPC decoding uses the 3GPP NR LDPC base graph 1 with a layered Min-Sum decoder for up to 25 iterations. Perfect timing and frequency synchronization are assumed. Each slot uses an independent channel realization (no fixed seed). All 16 scenarios use 200 slots per SNR point over the SNR range −2 to +18 dB in 2 dB steps. Latency measurements use 100 timed repetitions after 10 warmup repetitions; warmup runs are discarded to eliminate kernel-compilation and memory-allocation effects.

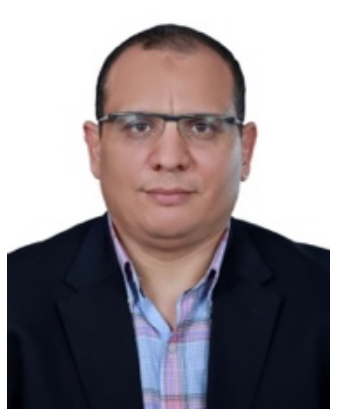

**AYMAN ELNASHAR** received his B.Sc., M.Sc., and Ph.D. degrees in Electrical Communications Engineering (1995, 1999, 2005) and holds an Executive MBA. He has more than 30 years of combined industry and academic experience spanning senior leadership roles, applied research, and teaching.

He is currently Vice President - Technology Strategy, Architecture & Innovation at e& UAE (etisalat), where he leads integrated strategies and innovation across mobile and fixed networks, IT, multi-cloud, and AI. He has spearheaded national and regional programs, including the region's first 3G, 4G, and 5G launches. His work also includes large-scale telco and IT cloud modernization, AI-first transformation, sovereign AI cloud initiatives, API and platform monetization, and end-to-end automation. In parallel, he serves as an Advisory Board Member at Lablabee, driving innovation and strategic oversight in technology education.

Dr. Elnashar previously held senior leadership roles at du (UAE), Mobily (KSA), and Orange (Egypt), where he contributed to the design and rollout of large-scale mobile, fixed, and cloud infrastructures. He also plays a prominent role in the academic and innovation ecosystem as an Industry Advisory Board Member at NYU Abu Dhabi, RIT Dubai, and AUS.

He has authored three Wiley reference books and an extensive body of peer-reviewed publications and industry whitepapers, spanning applied 6G/5G/4G/IoT system design and theoretical advances in adaptive detection and beamforming, digital signal processing, and neural networks. His research interests span wireless networks (6G/5G/SatCom/WiFi), AI/ML, AI-native networks, cloud-native systems, smart cities, and IoT, with a focus on translating advanced theory into practical, real-world ICT solutions.